\documentclass[runningheads]{llncs}
\usepackage[T1]{fontenc}
\usepackage{graphicx}
\usepackage{ifthen}
\usepackage{color}
\usepackage{amsmath,amssymb,amsfonts,array}
\newcolumntype{C}[1]{>{\centering\arraybackslash}m{#1}}
\usepackage{graphicx}
\usepackage{psfrag}
\begin{document}
\title{Permutation Entropy as a Conceptual Model to Analyse Brain Activity in Sleep}
\titlerunning{Permutation Entropy as a Model to Analyse Brain Activity in Sleep}
\author{Alexander Edthofer\inst{1}\orcidID{0000-0002-5669-705X} \and 
	Iris Feldhammer \inst{1} \and \\
	Thomas Fenzl \inst{2} \and
Andreas K\"orner\inst{1,*}\orcidID{0000-0001-7116-1707} \and \\
Matthias Kreuzer \inst{2}\orcidID{0000-0003-2472-3556}
}
\authorrunning{A. Edthofer et al.}
\institute{Institute of Analysis and Scientific Computing, TU Wien, \\ Wiedner Hauptstra\ss e 8, 1040 Vienna, Austria \and
	Department of Anaesthesiology and Intensive Care Medicine, \\ University Hospital Rechts der Isar, Technical University of Munich, \\ Ismaninger Stra\ss e 22, 81675 Munich, Germany \\
\textsuperscript{*}\email{andreas.koerner@tuwien.ac.at}
}
\maketitle            
\begin{abstract}
Sleep stage classification is a widely discussed topic, due to its importance in the diagnosis of sleep disorders, e.g. insomnia. Analysis of the brain activity during sleep is necessary to gain further insight into the processing that occurs in our brains. We want to use permutation entropy as a model for this analysis. Therefore, the signal processing in terms of electroencephalography is described. This results in a time discrete signal, that can be further processed by applying the method of permutation entropy, which is a modification of the Shannon entropy as a measure of information processing.
The method is applied to 18 data sets, nine electroencephalography measurements of patients suffering from insomnia and nine of people without a sleep disorder. A strong correlation between the permutation entropy value and the sleep stages was found during the simulation runs. The results are analysed and presented using boxplot diagrams of the permutation entropy over the sleep stages. Furthermore, it is investigated that there is a steady decrease in the value when the patient is in a deeper sleep. This suggests that the method is a good parameter for sleep stage classification. Finally, we propose an extension of the conceptual model to other pathological conditions and also to the analysis of brain activity during surgery. 

\keywords{permutation entropy \and sleep stage classification \and EEG monitoring.}
\end{abstract}

\section{Introduction}

Every day humans have to process a lot of information. This results in brain activity, which can be measured by electroencephalography (EEG). Plenty of techniques and methods for analysing the complex processes of our brains already exist. One of them is the permutation entropy (PE), which was introduced in \cite{bandt2002permutation}. It is a modification of the Shannon entropy, which is a specific measure of information processing, as it describes the rate at which information is produced when discrete information is modelled as a Markov process, see \cite{shannon1948mathematical}. Based on this, the PE is a complexity measure for time discrete signals as it maps tuples of a time series to a set of predefined patterns. The probability distribution of these is then used for the calculation of the measure.

The aim of this work is to apply the concept of the PE to sleep data and extract information about the sleep states of patients. The strong correlation between the entropy measure and the sleep stages was the fundamental motivation. On the one hand, high values of PE indicate noise-like behaviour and, therefore, wakefulness, where as on the other hand, low values indicate low activity and sleep. This allows the analysis of sleep behaviour and performance of sleep stage classification. The activity patterns are very differentiated during different sleep stages. Some areas are still very active, information is still being processed. Due to the great importance of sleep, the topic is widely discussed using a variety of methods \cite{kirsch2012entropy,mateos2017measure,zhang2018efficient,faust2018deep,faust2019review}. The long-term aim of this work is to have an open source parameter using PE that allows anaesthetists to monitor the level of a patient's coma during surgery, as was already suggested in \cite{olofsen2008permutation}.

Before applying the method, we first introduce the signal processing of brain activity and how to obtain a time discrete signal \cite{hoffmann2007geraete,sanei2013eeg}. It is described how different technical subsystems interfere during the measurement and how this is smoothed again by filtering. Afterwards, the Shannon entropy is explained \cite{shannon1948mathematical} and, as a modification, the PE is introduced and discussed mathematically in detail in order to have a robust model, following \cite{bandt2002permutation,berger2017permutation,zanin2012permutation}. 

Next, in section \ref{sleepstageclassification}, we consider 18 data sets of the Cyclic Alternating Pattern (CAP) sleep data base \cite{terzano2001atlas}. As open source data on this topic is already available, we will determine the PE with the data and simulations to test our conceptual model. Reproducibility is given as well. In \cite{mateos2017measure} the method is used as well, but only performed on three patients, so we want to take a look at a larger sample size and also consider patients suffering from a sleep disorder. The processing of the EEG signals of the patients is described. Afterwards, the results of the application of our method are presented. The correlation between the PE and the sleep stages is calculated and the sleep stage classifications are presented in boxplot diagrams.

\section{Signal Processing of Continuous Brain Activity into a Time Discrete Signal} \label{signal}

The human brain responds with an electrical impulse to every stimulus it perceives, whether by sight, touch, taste, smell or hearing. This signal is transmitted from the peripheral nervous system to the brain, where the information is processed. This processing is done by changing the electrical membrane properties, i.e. the ionic conductivity, see \cite{cooper1974elektro}. The measurement of this brain activity is of great importance to neurologists in the diagnosis of diseases, but also for anaesthetists in monitoring during surgery. This measurement is called EEG.

Several things are needed to measure an EEG signal, as it is stated in \cite{hoffmann2007geraete}. First, there are electrodes, which are the connection between the tissue and the technical system. Usually at least three electrodes are used for an EEG during a surgery, when it is done for diagnostic reasons there are more in use. In order to compare results and have reproducibility, the placement of the electrodes is standardised with the ten-twenty system. It is not rigid but dynamic with specifications given in percentages, so that the application also works for smaller heads, such as infants or children have. The reference for the voltage is either the average of the value of all electrodes used, if it is a multichannel EEG, or there is a reference and a ground electrode in addition to the electrode for the measurement.

With the measurement done by the electrodes one can proceed with signal processing. An amplifier increases the strength of the signal, since the deviating value of the reference is usually small. Further signal processing consists of the detection of artefacts. There are biological ones produced by the patient, such as movement of the patient, sweating or twitching of muscles. Other artefacts can be caused by the technical equipment or instruments, e.g. the electrodes, the amplifier and all other used components. Hence, it is necessary to apply highpass and lowpass filtering to the signal. On the one side, the highpass filter eliminates the low-frequent disturbances, which usually consist of the biological artefacts, on the other side, the lowpass filter neglects the high-frequent noise that usually occurs when using technical instruments or electromyography (EMG), which measures muscle activity.

\begin{figure}[b!]
	\includegraphics[width=\textwidth]{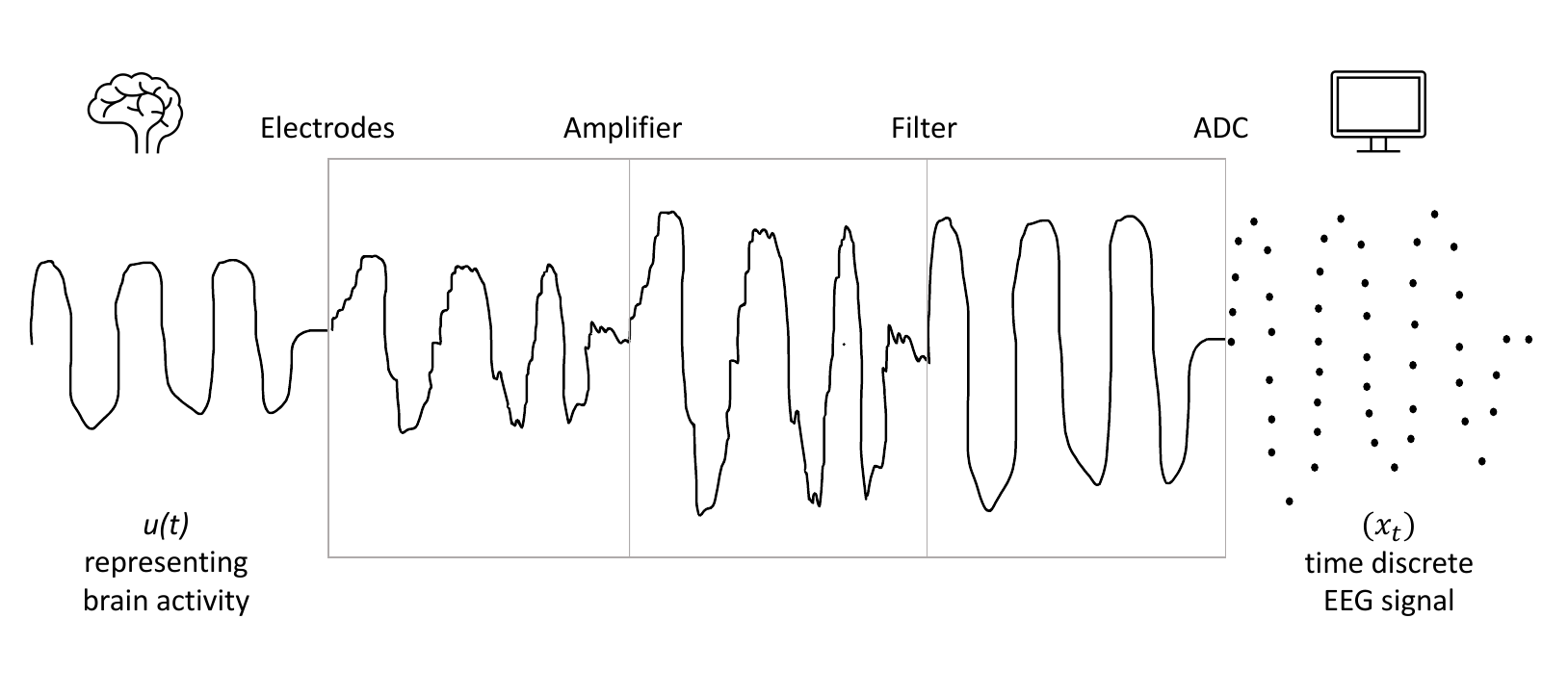}
	\caption{Schematic diagram of the signal processing of continous brain activity to get a time discrete signal.} \label{signalproc}
\end{figure}

With the signal processing also a discretisation goes hand in hand. This is done after filtering by a multichannel analogue-to-digital converter (ADC). To determine the correct sampling frequency, it is necessary to take a look at the frequency of human brain waves. They are classified in four categories, according to their frequency range: $\beta$ (13-30 Hz), $\alpha$ (8-13 Hz), $\theta$ (4-8 Hz), $\delta$ (0.5-4 Hz). Each of them appears at different stages of age or consciousness. Due to the Nyquist theorem \cite{shannon1949communication}, which states that a sampling rate of two times the maximal frequency is required, a sampling rate of at least 200 samples/s should suffice \cite{sanei2013eeg}. After the discretisation, further filtering can be done.

In Fig. \ref{signalproc} a schematic diagram of the whole process is shown. We now have a time discrete signal that represents our brain activity and can be further processed.

\section{Entropy as a Measure of Information}

After recording an EEG, we want to analyse it. Therefore, we want to calculate the PE value of the time discrete signal and see if it tells us anything about the brain activity. First, a few words about entropy in general. It has various applications but its origin is in physics, according to \cite{titz2013entropie}. There it is an important term in the field of thermodynamics. From there, entropy has spread to other areas of physics, from mechanics to astronomy. It has also found its way into chemistry as well as social sciences. The different fields of application interpret the term entropy in their own way, as \cite{popovic2017researchers} states. We will now look at its application in information theory, as Claude Shannon described it in his 1948 article "A Mathematical Theory of Communication" \cite{shannon1948mathematical}. 

The so-called Shannon entropy should describe the rate at which information is produced if discrete information is modelled as a Markov process. This means that we have $n$ possible events with certain probabilities $p_1, ..., p_n$. The Shannon entropy $H = H(p_1,...,p_n)$ is "a measure of how much 'choice' is involved in the selection of the event or of how uncertain we are of the outcome?", see \cite{shannon1948mathematical}, p. 10. It is derived from the assumptions: 
\begin{enumerate}
	\item $H$ is continuous in its arguments $p_j$.
	\item $H$ is monotonously increasing in $n$ if $p_j=\frac{1}{n}$ for all $j=1,...,n$.
	\item If there are three possible outcomes, there is a decomposition from three to two by adding another level of choice. 
\end{enumerate}

\noindent
With these assumptions and $K \in {\rm{I\!R^+}}$, Shannon showed in \cite{shannon1948mathematical} 
\begin{align}\label{shannon}
	H(p_1,...,p_n) = - K \sum_{j=1}^{n} p_j \log p_j.
\end{align}

\noindent
A modification of the Shannon entropy, the PE, was introduced by Christoph Bandt and Bernd Pompe in 2002 \cite{bandt2002permutation}, which will be the focus of this work. In the next section, the PE will be defined and discussed in detail.

\section{The Method of Permutation Entropy}

In this section, the PE, as Christoph Bandt and Bernd Pompe introduced in 2002 \cite{bandt2002permutation}, is discussed in more detail. It is a modification of the Shannon entropy as defined in equation (\ref{shannon}). It is a statistical measure and describes the complexity of a time discrete series. We orient the notation and definition according to \cite{berger2017permutation,zanin2012permutation}. For a given time series $(x_t)_{t=1,...,n}$ of length $n$ we can choose an order $m$ and a time delay $\tau$. The order specifies how long the investigated tuples of the time series will be and later we will see that it has a huge impact on the accuracy of the method. The time delay tells us the index distance between two values in a tuple. For $\tau =1$ we only have neighbouring values of the time series, for $\tau > 1$ this is no longer the case. If two values of the time sequence are equal $x_i = x_j$, for $i \neq j$, we neglect the value or add some random noise. Since we want to calculate the permutation of the time sequence, we divide in $k := n-(m-1) \tau$ tuples of length $m$. The mapping rule is 
\begin{align*}
	(x_1,...,x_n) \mapsto ((x_1, &x_{1+\tau},x_{1+2\tau}, ..., x_{1+(m-1)\tau}), \\ 
	(x_2,x_{2+\tau},x_{2+2\tau}&, ..., x_{2+(m-1)\tau}), ..., (x_k,x_{k+\tau},x_{k+2\tau}, ..., x_{k+(m-1)\tau})). \label{tuple}
\end{align*}

\noindent
Next, we take one tuple starting with the value $x_j$ and order the values from smallest to largest value. For numbers $r_1, r_2,...,r_m \in \{1,2,...,m\}$, where $r_1$ is associated with the first value of the tuple, which is $x_j$, $r_2$ is associated with the second value $x_{j+\tau}$, and so on, such that it holds
\begin{equation*}\label{order}
	x_{j+(r_s-1)\tau} < ... < x_{j+(r_1-1)\tau} < ... < x_{j+(r_{\tilde{s}}-1)\tau}.
\end{equation*}

\noindent
Hence, $r_s$ is the number, such that $x_{j+(r_s-1)\tau}$ is the lowest value of the whole tuple and $r_{\tilde{s}}$ such that $x_{j+(r_{\tilde{s}}-1)\tau}$ is the highest. The pattern $(r_1) ... (r_m)$ is a permutation $\omega_j$ of the pattern $(1)...(m)$. The set of all patterns is defined as $\Omega_m:=\{\omega | \omega \, \text{is a permutation of} \, (1)(2)...(m)\}$. The number of possible permutations of natural positive numbers up to $m$ is $m!$. Consequently, the cardinality of the set $\Omega_m$ is $|\Omega_m| = m!$. As an example, $\omega_1 = (1)(2)...(m-1)(m)$ and $\omega_{m!}=(m)(m-1)...(2)(1)$. By using the notation of brackets $(\, )$ around each digit, one can also use patterns of order greater than 9.

\begin{figure}
	\includegraphics[width=\textwidth]{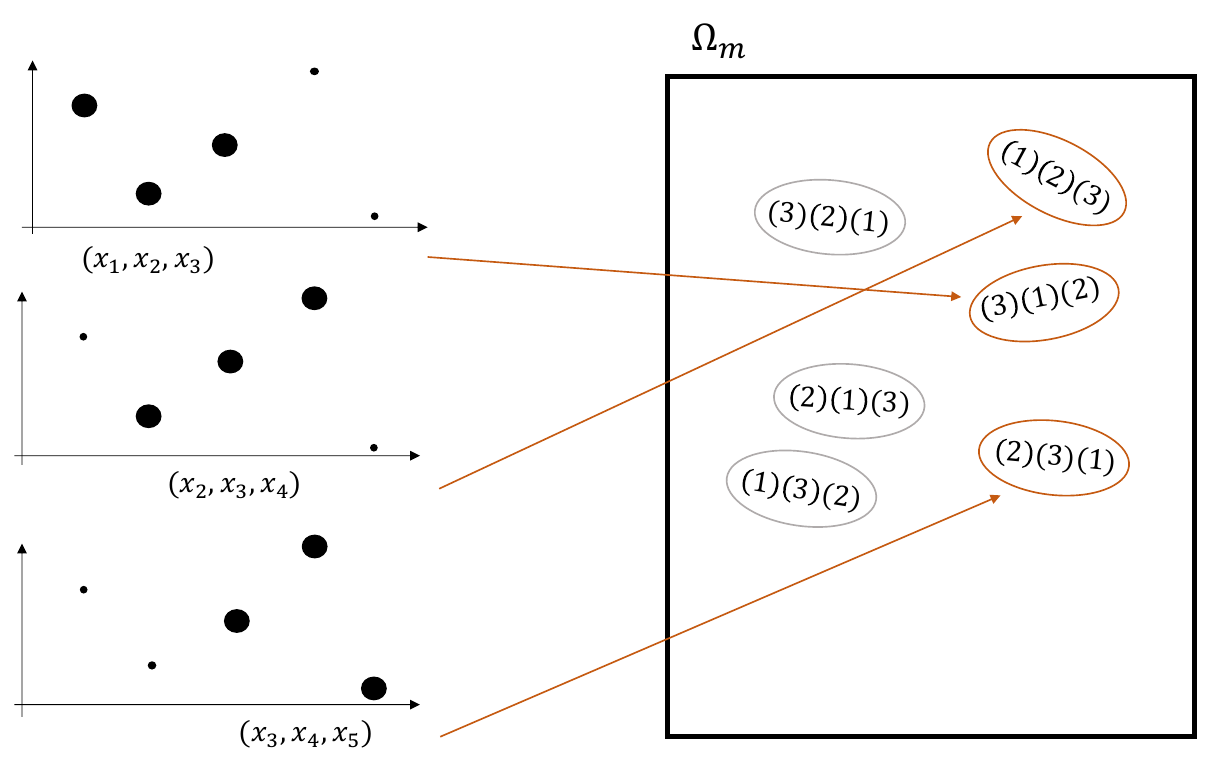}
	\caption{Schematic diagram of the correspondence between the tuples of the time discrete signal and its permutation type in $\Omega_m$ for $m=3$ and $\tau=1$.} \label{correspondence}
\end{figure}

Now the idea of the PE is to determine the permutation type $\omega_j$ of each tuple, so we get $k$ patterns, as it is done in Fig. \ref{correspondence} for $m=3$ and $\tau =1$. The probability distribution is then calculated and the probability of each type is defined as $p_j := P(\omega_j)$, where
\begin{align*}
	P(\omega_j) = \frac{\#\{i|1 \leq i \leq k, (x_i, ..., x_{i+(m-1)\tau}) \, \text{ has type } \omega_j\}}{k}.
\end{align*}

\noindent
According to the Shannon entropy, the PE is defined as 
\begin{align*}
	H(m) := H(p_1,...,p_{m!}) = - \sum_{j=1}^{m!} p_j \log p_j.
\end{align*}

\noindent
The base of the logarithm is 2. If the probability $p_j$ were equal to 0, the summand would not be defined. Therefore, we will take the limit of such a value towards 0 and, with the rule of de l'Hospital, see that
\begin{align*}
\lim_{p_j \rightarrow 0} p_j \log p_j = 0.
\end{align*}

\noindent
Hence, we can neglect such a summand in the overall sum. According to \cite{shannon1948mathematical} the range of $H$ is $[0,\log(m!)]$. The minimal value 0 is obtained, if there is a $\tilde{j}$ such that $p_{\tilde{j}} = 1$ and correspondingly $p_j = 0$ for all $j \neq \tilde{j}$. Then 
\begin{align*}
	H(m) = - \sum_{j=1}^{m!} p_j \log p_j = - p_{\tilde{j}} \log p_{\tilde{j}} = - 1 \cdot 0 = 0.
\end{align*}

\noindent
The maximal value $\log(m!)$ is obtained if there is an equal distribution over all patterns such that $p_j = \frac{1}{m!}$ for all $j \in {1, ... , m!}$.
\begin{align*}
	H(m) = - \sum_{j=1}^{m!} p_j \log p_j = - m! \frac{1}{m!} \log\left(\frac{1}{m!}\right) = - \left(\log 1 - \log(m!)\right)= \log (m!)
\end{align*}

\noindent
Consequently, we can define a normalised PE by dividing by $\log(m!)$, according to
\begin{align}\label{normalPE}
	\hat{H}(m) = - \frac{1}{\log(m!)} \sum_{j=1}^{m!} p_j \log p_j. 
\end{align}

\noindent
This value allows us to compare the results for different orders as well as to define certain ranges in an application. By comparing with the extremal values, which were already checked above, one can easily verify that if a pattern has probability $p_{\tilde{j}} = 1$ again then $\hat{H}(m) = 0$ and if we have an equal probability distribution of $p_j = \frac{1}{m!}$ then $\hat{H}(m) = 1$. Hence, large values of the normalised PE are a sign of high complexity.

Now we will take a look at the various applications, where the PE is already implemented. As Yan et al. explain in \cite{yan2012permutation}, the value can characterise the status of rotary machines. Our field of interest is the application to the analysis of an EEG. It has already been used quite often in research, for example by Olofsen et al. in \cite{olofsen2008permutation}. They describe the effect of anaesthetics on the human brain activity by applying the PE to the time series obtained by the signal processing of an EEG. In this work, we would like to take up this point and present with sleep stage classification another application of the PE based on the analysis of EEG data. From now on the definition of the PE from equation (\ref{normalPE}) is used.

\section{Sleep-Stage Classification using Permutation Entropy}\label{sleepstageclassification}

For sleep analysis it is important to know the specific sleep states, i.e. to be able to "measure" and determine them. There are two main categories of sleep: rapid eye movement sleep (REM) and nonrapid eye movement sleep (NREM). The latter is further divided into sleep stages 1 to 4. Stage 1 is drowsiness, when someone is in between being awake and asleep. During a sleep cycle, stage 2 takes around 40 to 50 \% of the time, which is when brain waves slow down. In stage 3, $\delta$-waves, which were mentioned in section \ref{signal}, begin to occur. In stage 4 these waves become dominant. Stages 2, 3 and 4 are called light sleep, deep sleep and very deep sleep, respectively, see \cite{sanei2013eeg}. Including the stage of being awake we have a total of 6 stages that can be classified. This goes back to the year 1968, published by Rechtschaffen and Kales \cite{rechtschaffen1968manual}. In 2007, the American Academy of Sleep Medicine introduced a new classification by merging stage 3 and stage 4, which is internationally recognised and used, see \cite{iber2007aasm}. 
  
Due to its importance, the analysis of sleep is an already widely discussed topic \cite{kirsch2012entropy,mateos2017measure,zhang2018efficient,faust2018deep,faust2019review}. The first three publications apply methods using entropy to an EEG for further investigation, but only \cite{mateos2017measure} uses PE. In the latter publication the method is only applied to three patients, so we wanted to take a look at a larger sample size and look at patients suffering from a sleep disorder as well.

\begin{figure}
	\includegraphics[width=\textwidth, height=0.4 \textheight]{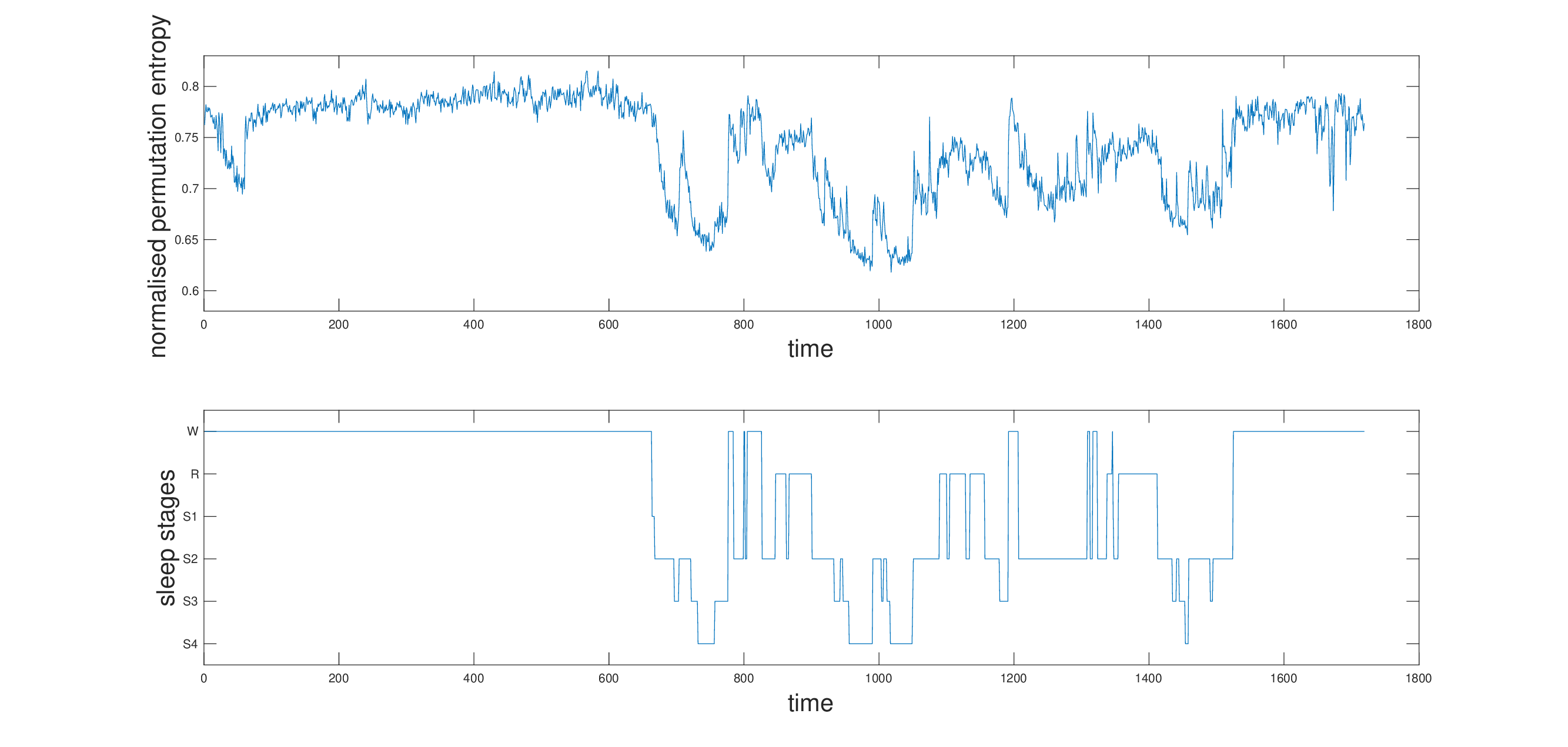}
	\caption{Course of the normalised PE and the sleep stages over time using the example of the patient ins5.} \label{timecourse}
\end{figure}

As a motivating example for the correlation between the normalised PE and the sleep stage classification take a look at Fig. \ref{timecourse}. There it is illustrated that the PE and sleep stages have a similar course when watched over time.

The normalised PE will now function as the underlying modelling method to analyse the sleep stages from the EEG via simulation. We used a part of the open source CAP sleep database \cite{terzano2001atlas} to apply this classification model. Since it was published in 2001, the sleep stage classification by Rechtschaffen and Kales was used. The database is a collection of 108 recordings of people suffering from a sleep disorder, but also including a control group, which are people without any sleep disorder. Every data set included, among other things, at least three EEG signals. We selected 18 patients. Nine of them are n1 to n9, people without any sleep-related condition, and the other nine are ins1 to ins9, patients suffering from insomnia, a sleep disorder. The data set also contained information about the sleep stages of the patients, which are given in a 30 second intervals. Because of this 30 second window we also split the EEG into 30 second windows. The implementation was done in MATLAB\textsuperscript{\copyright}, Version 2022b. It is based on the code of S. Berger, published in \cite{berger2017permutation,berger2019teaching}, but adapted and extended to the specific needs of this project. The given EEG is resampled to a sampling frequency of 200 Hz and afterwards low-pass filtered at 30 Hz again. The measured data is then partitioned into windows of 6000 data points, corresponding to 30 seconds of signal duration. Then, the normalised PE is calculated of each of the windows, which associates an entropy value to every given sleep stage classification. We will take a closer look at the correlation between the two. For the calculation, we assigned each sleep stage a value, as displayed in Table \ref{ssv}. The stages are abbreviated as W for wakefulness, R for REM sleep and S1 to S4 for the sleep stages 1 to 4, respectively. Nowadays the non-REM sleep is abbreviated as N instead of S and there are only three stages of this type of sleep.

\begin{table}
	\caption{Abbreviations and values of each sleep stage.}\label{ssv}
	\begin{tabular}{|c||C{1.6 cm}|C{1.6 cm}|C{1.6 cm}|C{1.6 cm}|C{1.6 cm}|C{1.6 cm}|}
	\hline
	Sleep stage & stage 4 & stage 3 & stage 2 & stage 1 & REM sleep & awake \\
	\hline
	\hline
	Abbreviation & S4 &  S3 & S2 & S1 & R & W \\
	\hline
	Value & 0 & 1 & 2 & 3 & 4 & 5 \\
	\hline
	\end{tabular}
\end{table}

\noindent
First we determined which combination of order and time delay gives a good correlation measure. A small investigation for the patients ins1-ins4 and n1-n4 showed that an order of $m=3$ and a time delay of $\tau=1$ give the best correlation. There were also good values for order $m=7$ as well as for time delay of $\tau=3$ (for orders $m=3$ and $m=7$) but a higher time delay led to a lower correlation value. This is already indicated in \cite{popov2013permutation}, where it is described that higher time-delays behave the same as lower sampling frequencies.

Each patient had multiple measurement electrodes, so we had to decide which electrode to take the EEG from in order to apply the method. The measurement of the electrodes Fp2-F4 was taken from all patients, except n6, n7 and n9, which showed a strong correlation in all of them. For the patients n6, n7 and n9 we chose the measurement of the electrodes C3-A2 because it also showed a strong correlation.

The data set provided the sleep stage classification only in a certain time range, hence we had to limit the measured EEG to the corresponding time range. There was also a difference in the recording time between the patients, which resulted in a different amount of values. The final range was between a measured time of 6 hours 6 minutes and 14 hours 20 minutes, corresponding to 732 and 1719 values respectively. All simulations and classifications were performed and evaluated in MATLAB\textsuperscript{\copyright}.

\section{Results}

\begin{figure}
	\includegraphics[width=\textwidth, height=0.43\textheight]{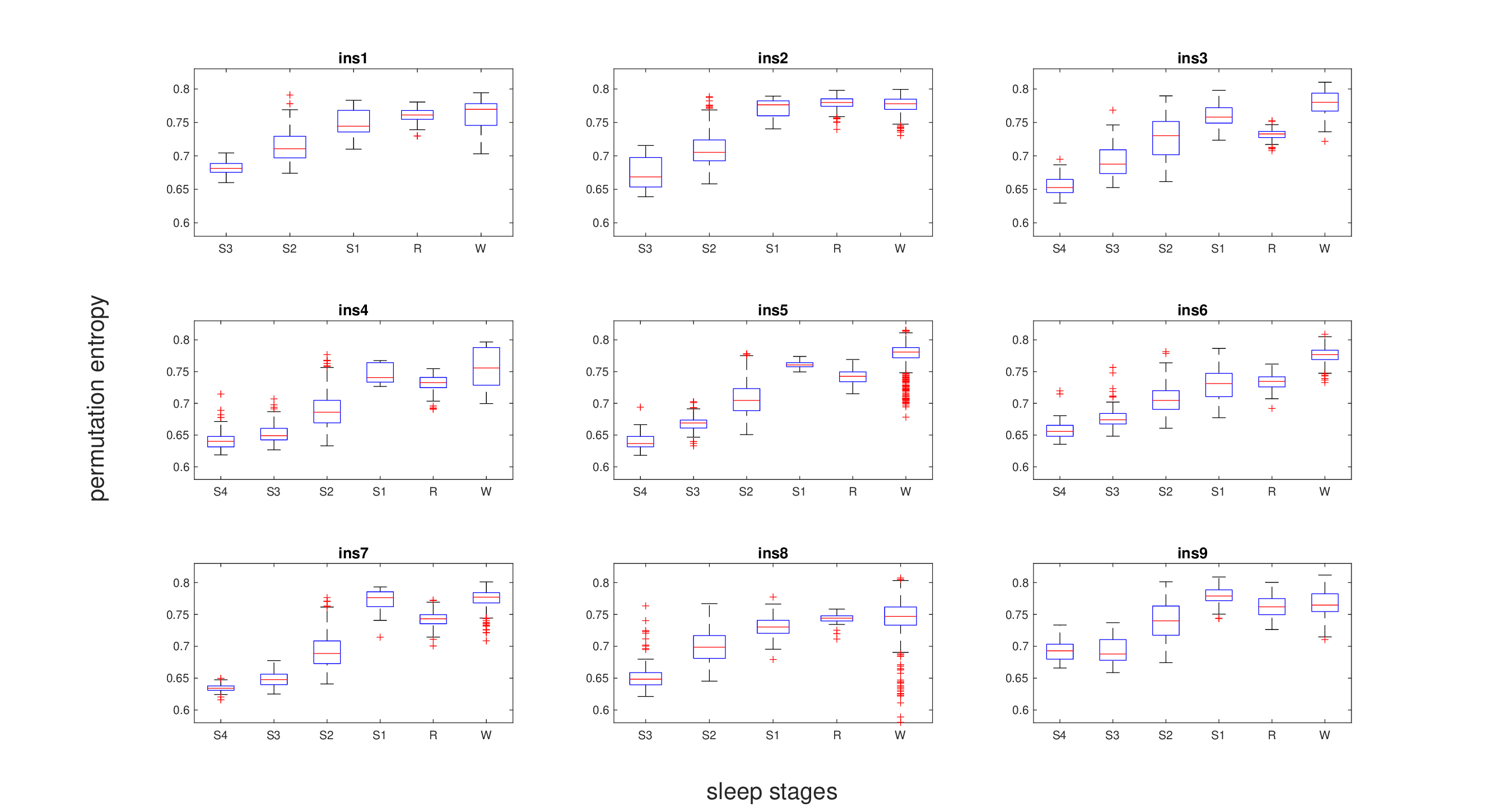}
	\caption{Boxplots of the normalised PE over the sleep stages of each of the patients suffering from insomnia.} \label{insomnia}
\end{figure}

First, the patients suffering from insomnia are considered and the normalised PE is computed. Then, also its correlation with the assigned values to the sleep stages, which are listed in Table \ref{ssv}, is calculated and shown in Table \ref{corr1}. In order to have a closer look at the sleep stage classification, boxplot diagrams of the normalised PE for each of the sleep stages of each patient were generated with the pre-implemented function of MATLAB\textsuperscript{\copyright}, see Fig. \ref{insomnia}. 

\begin{table}
	\caption{Correlation between normalised PE and sleep stages for data of the patients suffering from insomnia.}\label{corr1}
	\begin{tabular}{|c||c|c|c|c|c|c|c|c|c|}
		\hline
		Patient& ins1 &  ins2 & ins3 & ins4 & ins5 & ins6 & ins7 & ins8 & ins9 \\
		\hline
		\hline
		Correlation& 78.62\% & 87.43\% & 80.79\% & 82.25\% & 91.11\% & 92.36\% & 91.23\% & 70.45\% & 62.39\% \\
		\hline
	\end{tabular}
\end{table} 

\noindent
Table \ref{corr1} shows a strong correlation for all nine data sets. The only one with a correlation below 70\% correlation is the one of patient ins9. One reason to explain this is due to the large overlap of the values of the sleep stages S4 and S3, as well as S1, R and W.  Looking more closely at the boxplot diagrams one can see the respective ranges of the PE of the sleep stages of each patient. Single outliners are marked with a red "+". {They are determined by the default settings of the boxplot function of MATLAB\textsuperscript{\copyright} where the whiskers are set to 1.5 times the interquartile range. Thus, any data point that is more than 1.5 times the interquartile range away from its quartile is an outliner.} 

\noindent
Furthermore, it is visible that the patients ins1, ins2 and ins8 never reach sleep stage 4. Since S3 and S4 are merged in the nowadays used sleep stage classification by the American Academy of Sleep Medicine, this fact is negligible.

\begin{figure}
	\includegraphics[width=\textwidth, height=0.43\textheight]{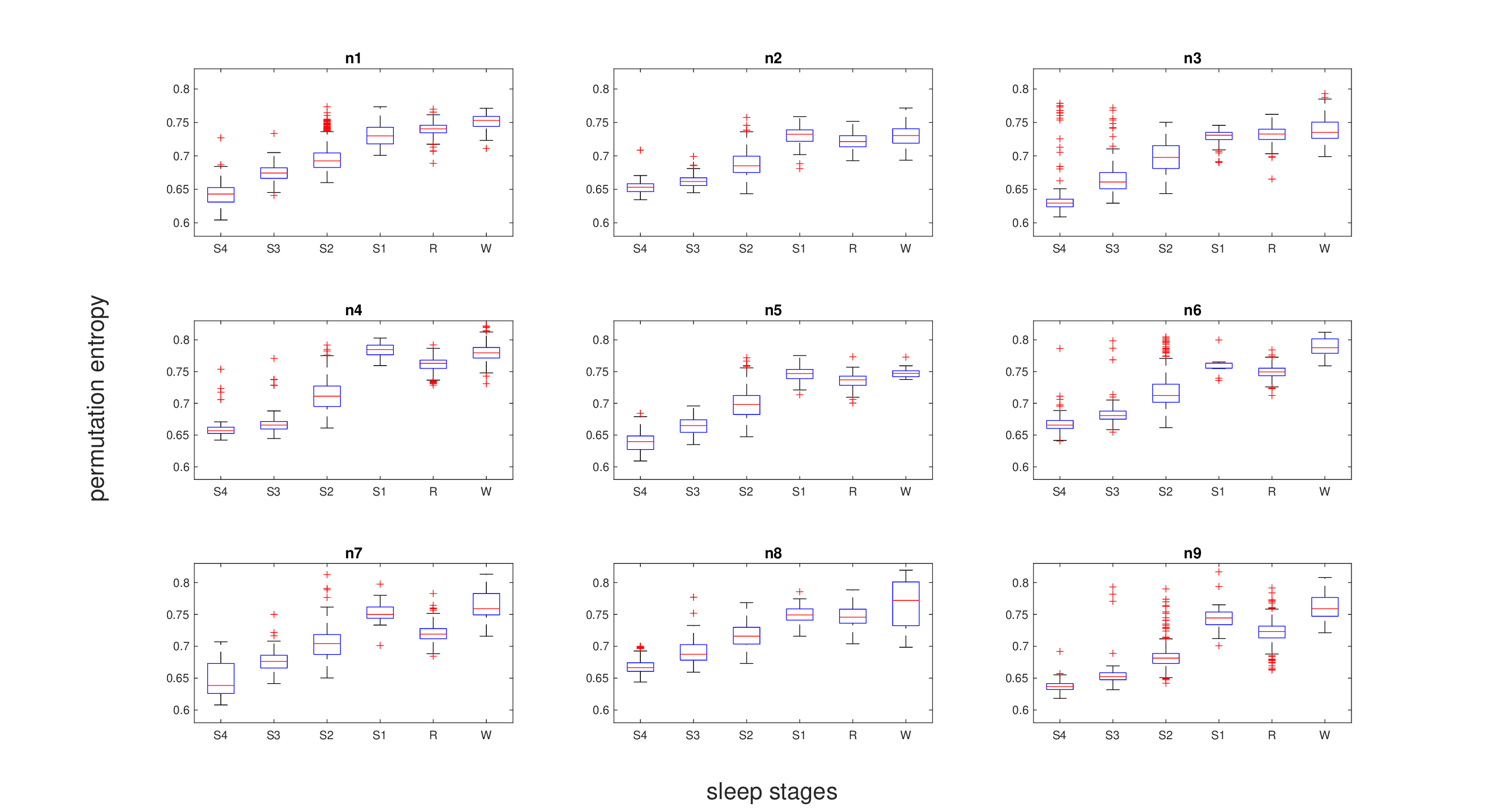}
	\caption{Boxplots of the normalised PE over the sleep stages of each of the patients without a sleep disorder.} \label{normal}
\end{figure}

For comparison, we now look at Table \ref{corr2} and Fig \ref{normal}, where the correlation value and the boxplots for the patients without a sleep disorder are shown. 
This time we have a very strong correlation for all studied subjects.

\begin{table}
	\caption{Correlation between normalised PE and sleep stages for data of the patients without a sleep disorder.}\label{corr2}
	\begin{tabular}{|c||c|c|c|c|c|c|c|c|c|}
		\hline
		Patient& n1 &  n2 & n3 & n4 & n5 & n6 & n7 & n8 & n9 \\
		\hline
		\hline
		Correlation& 87.81\% & 80.96\% & 79.60\% & 87.86\% & 86.32\% & 81.49\% & 77.53\% & 81.21\% & 86.44\% \\
		\hline
	\end{tabular}
\end{table}

\noindent
Overall there is a difference between each range of the PE value, but all are between 0.58 and 0.83. The correlation coefficients are also similar in both groups, with the exception of ins9. 
What all studies have in common is the steady increase in the PE value as the patients become more awake, i.e. when comparing sleep stages S4-S1 and awake. This applies to both groups, people with and without a sleep disorder. This was also indicated in \cite{mateos2017measure}, who showed it for a sample size of three. REM sleep mostly ranges somewhere between S1 and S2. 

The range of S2 is also the broadest one. This can be explained by the fact that it describes the phase of light sleep in which we spend 40 to 50 \% of the time of one sleep cycle, as already mentioned in section \ref{sleepstageclassification}. 

\section{Discussion}

In our work we used the PE as a modelling method for analysing brain activity. The signal processing in terms of an EEG was described and the method itself was specified in every mathematical detail. Furthermore, the importance of sleep stage classification was motivated and then also performed with the normalised PE for 18 data sets of patients suffering from the sleep disorder insomnia and people without any sleep-related sickness. Within the investigation we showed that there is a strong correlation between the PE and sleep stages for all of the patients. Also a connection between deeper sleep and lower value of the normalised PE was shown, suggesting that it is a good parameter for sleep stage classification. 

We looked at one EEG signal per patient, which is a current limitation that we will be working with. Whenever possible, we chose the measured signal Fp2-F4, which had the best correlation value on average among all data sets. Here, different electrodes or the combination of more signals may give a better insight into the classification. Moreover, we applied the method with order $m=3$ and with time delay $\tau = 1$, as it gave the best results on average. A detailed study of changing these parameters is not done in this work, but will be considered in the future. Regarding sleep disorders, we only selected patients suffering from insomnia as the most common sleep disorder \cite{someren2021brain}. Perhaps interesting results could be obtained by looking at sleep stage classification of different disorders in consultation with experts in this field.

Other variants of spectral entropy measures work well for sleep stage classification. Examples are the Walsh spectral entropy and Haar spectral entropy \cite{kirsch2012entropy} and the fuzzy measure entropy \cite{zhang2018efficient}. In \cite{faust2018deep} a different approach is proposed. Deep learning methods for physiological signals are described and it is stated that this method improves more and more  over time as more data becomes available. Another perturbation, the multivariate multi-scale weighted permutation entropy, is also proposed in the diagnosis of Alzheimer's disease in \cite{deng2017multivariate}.

Sleep scoring systems have already been developed for the electrocardiogram, electroencephalogram and electrooculogram. All of these signals contain information regarding the sleep stages \cite{faust2019review}. A combination of these signals is likely to give a better insight. This indicates that adding a new parameter, which could be as simple as peripheral pulse or blood pressure, will lead to a better classification. With several measured parameters the stages can then be found by clustering the data points.

Sleep stage classification was the first step in a larger project proposal. It is planned to apply the concept of PE to the analysis of the brain activity during anaesthesia in a surgery.. This should allow the anaesthetist to control the level of a patient's loss of consciousness, depending on the PE, among other parameters. In order to find the optimal amount of anaesthesia, i.e., not to give too much or too little anaesthetics, further research in this area is important \cite{olofsen2008permutation}.

In our work, we have learnt some lessons and picked up aspects that we want to investigate further. As the results for the classification of sleep stages met our initial expectations, we intend to continue our work and apply it in the field of anaesthesia for the classification of stages of loss of consciousness during surgery.

%
%
%
%

\end{document}